\documentstyle[12pt,moriond]{article}
\include{psfig}
\begin{document}
\heading{MODELLING HIGH-Z GALAXIES\\ FROM THE FAR-UV \\TO THE FAR-IR}

\author{J. E. G. Devriendt $^{1}$, 
B. Guiderdoni $^{1}$, S. K. Sethi $^{1}$} 
{$^{1}$ Institut d'Astrophysique de Paris, Paris, France.}
\\

\begin{moriondabstract}
In this paper, we report on a first estimate of the 
contribution of 
galaxies to the diffuse extragalactic background from the
far-UV to the submm, based on semi--analytic models of 
galaxy formation and evolution.
We conclude that the global multi--wavelength picture 
seems to be consistent provided a quite 
important fraction of star--formation be hidden in 
dust--enshrouded systems at intermediate and 
high--redshift. 
We show that, according to such models,
galaxies cannot stand as important contributors to the 
background hydrogen-ionizing flux at high-redshift unless neutral
hydrogen absorption sites are clumpy and uncorrelated with star 
forming regions.
We briefly discuss the robustness of such a result. 
\end{moriondabstract}

\section{Introduction}

In the paradigm of hierarchical structure formation, 
small dwarf--like objects are expected to form first
and dominate the global population of objects at 
high--redshift. To study the cosmological
influence of such galaxies as we wish to do here, 
one needs an accurate modelling of their energy budget, the
key factors to obtain a correct estimate of fluxes being star 
formation rates and absorption. 
This is the reason why, in this work, we include 
a multi--wavelength approach
in simple semi--analytic models of galaxy formation and
evolution.  

\section{A Short Guide to Semi-Analytical models}

Start from the power spectrum of linear fluctuations, $P(k)$, and
provided you pick a specific cosmology ($H_0$, $\Omega_0$, 
$\Lambda_0$) as
well as a candidate for dark matter (CDM, HDM, CHDM ...), you can
derive the number of
dark matter haloes of a given mass $M$ which collapse at a specific 
redshift $z$, using a Press-Schechter like prescription. Further assume the collapse to be described by a spherical 
model (TOP HAT), and you can then let the baryonic gas cool in the
potential wells of virialized dark matter haloes, watch it 
settle into 
disk--like structures and start forming stars. These stars interact with
the medium in which they formed, enriching it with metals and dust or
blowing it away. Giving reasonable prescriptions to describe these 
mechanisms enables one to compute statistical properties 
of galaxies,
say, their luminosity function $\phi(L)$. As demonstrated
for instance by \cite{kau}, such semi-analytic models can match the
optical data quite well. We have extended such models to the IR
window, in order to be able to handle the energy budget of galaxies 
correctly. In the following, we discuss results we obtained thanks to
this extension. We use a SCDM with
$H_0=50$ km s$^{-1}$ Mpc$^{-1}$, $\Omega_0=1$, $\Lambda_0=0$, 
but we would like to draw the attention on the
fact that the key parameter is the star
formation efficiency, and we distinguish two classes of
models in the following. Models where galaxies can form 
stars very 
efficiently at some point in their history will be referred to as
``starburst'' models, and are opposed to ``quiescent'' 
models where 
galaxies form their stars with an efficiency typical of
the Galaxy.

%\begin{center}
%{\bf Table 1.} Margins for two sizes of paper.
%\end{center}
%\begin{center}
%\begin{tabular}{|l|l|l|}
%\hline Margins & Type A4 & US Letter \\ \hline
%     &        &     \\
%Left & 2 cm & 2.3 cm \\ Right & 2 cm & 2.3 cm \\ Top & 2.4 cm & 1.5 cm \\
%Bottom & 2.4 cm & 1.5 cm \\ \hline
%\end{tabular}
%\end{center}
%\medskip

\section{Diffuse Extragalactic Background Light: A First Estimate}

Computing the contribution
of galaxies to the extragalactic background light
in wavelength bands ranging from the optical 
to the submm is the first step in the process of 
checking that the
energy budget of galaxies is correctly 
handled by the models. 
\begin{figure}[h]
\centerline{\psfig{figure=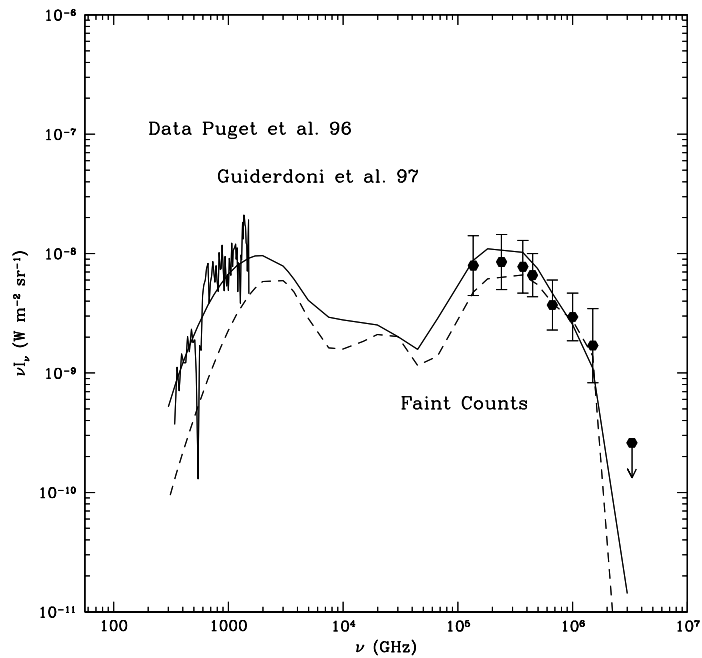,width=0.55\textwidth}}
{\caption{ This figure represents the contribution of galaxies
to the diffuse extragalatic background light at $z\simeq0$. 
Data is taken from \cite{gui1}, \cite{pug}, \cite{wil}.
The solid line is the predicted contribution from the 
semi-analytic
models with a ``burst--like'' star formation rate at 
intermediate and high--redshift.
The dashed line is the background flux predicted by the same models
with a ``quiescent'' star formation rate (see text). 
For details see \cite{jd2}, \cite{gui2}.}
  \label{back}}
\end{figure}
This is due to the
fact that this UV to submm background gives a fair estimate
of the amount of energy released by cosmic star formation 
at all epochs in the universe.
Recent observations (see \cite{gui1}) 
have enabled a detection of this background in the IR/submm
 and the lower 
limits given by faint counts in the optical and near IR both from the
ground and from space with HST are very close to a detection of the 
optical background (see \cite{poz}).
As shown in Fig. \ref{back}, models with ``quiescent'' star formation
rates (meaning a star formation history which varies
slowly as in our own Milky Way) do not reproduce the detected IR/submm extragalactic
background flux. A burst--like behaviour for star formation
(as the one we will use to derive our estimates of the UV
background flux) is required
at intermediate and high--redshift if one wants to match
the amount of energy re-radiated in the IR/submm window.
This simply means that an important amount of star 
formation is likely to have
taken place in dust--enshrouded regions in the past.
 
\section{Evolution of the Lyman-limit background}

A generic prediction of the hierarchical
scheme of structure formation is that the building
blocks (galaxies) were smaller and denser in the past compared to what
they are now, and that big objects formed from the merging
of smaller subunits. The question we would like to address 
here is the following:
can such dwarf--like objects that, according
to these
models, populate the universe at high--redshift, be 
responsible for its re--ionization?
\begin{figure}[h] 
\centerline{\psfig{figure=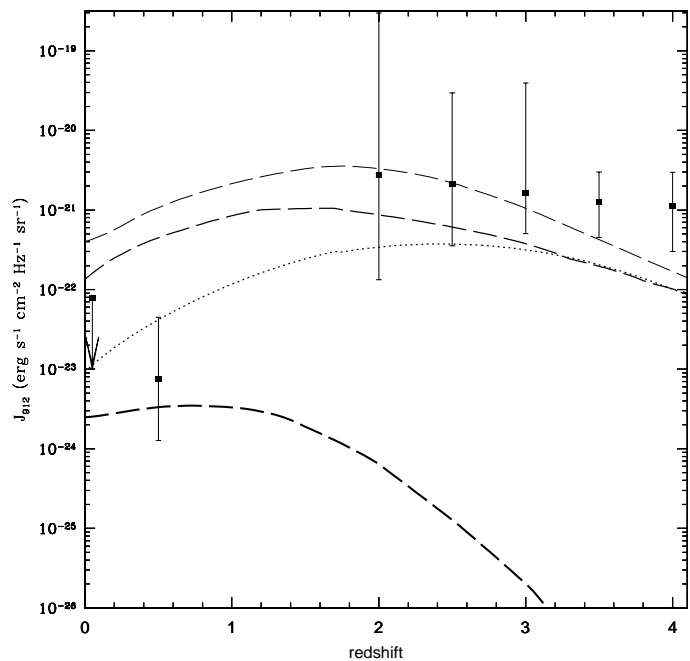,width=0.55\textwidth}}
\caption{ This plot shows the evolution of Lyman-limit flux with
redshift. The {\em dashes \/} of
increasing thickness correspond respectively to ``starburst''
models without dust
and HI absorption, with dust but without HI absorption, and  with both dust
and HI absorption. 
The {\em dotted \/} line  represents the contribution of quasars to  
the Lyman-limit background flux. 
For details see \cite{jd1}. Data are from \cite{coo}, 
\cite{kul}, \cite{vog}.
\label{ion}}
\end{figure}
In the previous section we derived what seemed to be a plausible
cosmic star formation history, but,
in order to answer this question, one needs to estimate
the crucial factor of HI absorption in both the ISM 
and the IGM (see \cite{jd1} for details). 
Then, one is naturally able to derive the time 
evolution of the 
Lyman-limit background from star--forming galaxies.
This is what is shown in Fig.~\ref{ion}. Two
conclusions can be drawn from this figure. 
First, galaxies produce the required amount 
of luminosity to explain the
observations.
Second, it is clear  
that, once dust and HI absorption are taken into account,
this contribution drops by
several orders of magnitude below that of quasars for 
nearly all redshifts but $z\simeq 0$. 
These results are mainly based on the 
assumption that
the UV flux seen by Madau and co-workers (see \cite{mad}) 
in the Hubble Deep Field has been correctly estimated. 
However, one would have to increase this flux by more 
than an order of magnitude and invoke a very clumpy 
distribution of HI in the ISM, 
to turn galaxies into important contributors to the Lyman--limit
flux. Hence we believe that the limits we derived for the
contribution of star--forming galaxies to the Lyman--limit
background flux are fairly robust. 
A more detailed discussion may be looked up in \cite{jd1}.

We conclude that in many respects, the picture that 
emerges in different windows from
the UV to the submm does not seem to be inconsistent.

%equation~(\ref{eq:curv}):
%\begin{equation}
%R_0=\frac{c}{H_0|\alpha_0|^{1/2}} {\rm~~with~~} \alpha_0=\Omega_0+\lambda_0-1
%\ .
%\label{eq:curv}
%\end{equation}

%
\vfill
\end{document}